\documentclass[aps,twocolumn,epsf,floats,pre]{revtex4-1}
\usepackage{graphics,graphicx,epsfig}
\usepackage{amssymb,color}
\usepackage{epsf,epstopdf,wrapfig}
\usepackage{amsmath}

\usepackage{amsmath}
\usepackage{amssymb}
\usepackage{bbold}
\usepackage{graphicx}

\def\vec#1{\mathbf#1}
\def\kk{\vec{k}}
\def\rr{\vec{r}}
\def\vv{\vec{v}}
\def\ww{\vec{w}}
\def\un{\check {\mathbf n}}
\def\vpar{\delta\vv_\parallel}
\def\vperp{\delta\vv_\perp}
\def\vp{\delta v_\perp}

\begin{document}

\title{Silent Flocks}

\author{Andrea Cavagna$^{a,b,c}$,  Irene Giardina$^{a,b,c}$, Tomas S. Grigera$^{d}$, Asja Jelic$^{a,b}$, Dov Levine$^{e,c}$, \\ 
Sriram Ramaswamy$^{f,g}$, and Massimiliano Viale$^{a,b}$}

\affiliation{$^a$ Istituto Sistemi Complessi, Consiglio Nazionale delle Ricerche, UOS Sapienza, 00185 Rome, Italy}
\affiliation{$^b$ Dipartimento di Fisica, Universit\`a\ Sapienza, 00185 Rome, Italy}
\affiliation{$^c$ Initiative for the Theoretical Sciences, The Graduate Center, CUNY, New York, NY 10016 USA}
\affiliation{$^d$ Instituto de Investigaciones Fisicoqu{\'\i}micas
  Te{\'o}ricas y Aplicadas (INIFTA) and Departamento de F{\'\i}sica,
  Facultad de Ciencias Exactas, Universidad Nacional de La Plata,
  c.c. 16, suc. 4, 1900 La Plata, Argentina} \affiliation{CONICET La
  Plata, Consejo Nacional de Investigaciones Cient{\'\i}ficas y
  T{\'e}cnicas, Argentina}
\affiliation{$^e$ Department of Physics, Technion-IIT, 32000 Haifa, Israel}
\affiliation{$^f$ TIFR Centre for Interdisciplinary Sciences, 21 Brundavan
Colony, Osman Sagar Road, Narsingi, Hyderabad 500 075, India}
\affiliation{$^g$ on leave from Department of Physics, Indian Institute of
Science, Bangalore 560 012, India}

\pacs{PACS numbers: 87.18.Gh, 05.65.+b, 47.54.-r, 87.18.Hf}

\begin{abstract}
Experiments find coherent information transfer through biological groups on length and time scales distinctly below those on which asymptotically correct hydrodynamic theories apply. We present here a new continuum theory of collective motion coupling the velocity and density fields of Toner and Tu to the inertial spin field recently introduced to describe information propagation in natural flocks of birds. The long-wavelength limit of the new equations reproduces Toner-Tu theory, while at shorter wavelengths (or, equivalently, smaller damping), spin fluctuations dominate over density fluctuations and second sound propagation of the kind observed in real flocks emerges. We study the dispersion relation of the new theory and find that when the speed of second sound is large, a gap sharply separates first from second sound modes. This gap implies the existence of ÔsilentÕ flocks, namely medium-sized systems across which neither first nor second sound can propagate.
\end{abstract}

\maketitle

%\pacs{PACS numbers: 87.18.Gh, 05.65.+b, 47.54.-r, 87.18.Hf}

%%% Intro

Models of self-propelled particles,  where dynamical equations for the individual velocities and positions are specified for each particle \cite{vicsek+al_95,gregoire+chate_04, hemelrijk_11,romanczuk_2012}, offer a microscopic description of a great variety of active matter systems \cite{vicsek_review, cavagna_giardina_14}, from granular materials \cite{granular}, to bacterial colonies \cite{chen} and animal groups \cite{couzin+krause_03}. By coarse-graining the microscopic models, it is possible to derive the hydrodynamic equations describing the dynamics of the velocity and density fields at large scales of length and time \cite{toner+tu_95, toner+al_05,toner+tu_98,toner+al_98, ramaswamy_review,marchetti_review}. The minimal model of collective motion is the Vicsek model \cite{vicsek+al_95}, whose continuous formulation has been provided by the elegant hydrodynamic theory of Toner and Tu \cite{toner+tu_95}.

The power of the hydrodynamic approach lies in the unambiguous choice of
variables -- only those whose timescale diverges in the limit of infinite system
size and zero wavenumber, $k \to 0$. Once these hydrodynamic variables,
determined by conservation laws and broken symmetries, are identified, the
theory is universal and independent of the details of the microscopic dynamics.
Natural systems, however, are often far from such limits and exhibit important
collective phenomena over medium scales. Flocks of birds are one such example.
These groups perform collective turns on so short a time scale that mutual
positions remain essentially the same and the coupling between density and
velocity fluctuations is very weak \cite{attanasi+al_14}. Hence, if a comparison
with biological data is in order, it is essential to go beyond the $k\to 0$
limit. This forces us to give up the gift of universality and to resort to
experimental  evidence in order to decide what is relevant and what is not in
the finite-size theory of collective motion. The scales involved still contain
a large number of birds, though, so a coarse-grained approach remains appropriate. 

It has been experimentally found in \cite{attanasi+al_14} that in order to describe information transfer through natural (and thus finite) flocks of birds, it is crucial to take into account inertial effects. In particular,  it is necessary to associate to the velocity a new quasi-conserved variable, the spin \cite{attanasi+al_14}. The spin is the generalized momentum generating the local rotations of the velocity field. Spin fluctuations transport across the flock the orientational information responsible for  the turn. The spin has an associated inertia, which is formally to the spin what standard mass is to linear momentum. The Inertial Spin Model (ISM) introduced in \cite{cavagna+al_14} is the microscopic self-propelled particle model coupling the dynamics of the velocity to that of the spin. It has been shown in  \cite{attanasi+al_14} and \cite{cavagna+al_14} that the ISM  quantitatively reproduces  the propagation of information across turning flocks of starlings. 

At very large scales (or, equivalently, for very large damping), the inertial effects in the ISM are irrelevant, but over finite scales they become essential. The nature of this crossover is not known, though, because the ISM has been analytically studied only in the short time limit of negligible density fluctuations \cite{cavagna+al_14}. Here we address this point by introducing a new set of continuous equations corresponding to the dynamical field description of the ISM.  As expected,  for $k\to 0$ the new field equations reproduce Toner and Tu's theory \cite{toner+tu_95}. However,  for finite $k$, and therefore finite length and time scales,  we find an entirely new and quite rich phenomenology. The most surprising result is the emergence of a gap in the dispersion relation, showing that propagation phenomena in flocks are possible only in certain size regimes.
§

%%%%%%%%%%%%%%%%%%%%%
%{\bf The new hydrodynamic equations.}
%%%%%%%%%%%%%%%%%%%%%

{\it The new theory.} 
We aim to write the hydrodynamic theory corresponding to the microscopic Inertial Spin Model (ISM) introduced in  \cite{cavagna+al_14}. To do this we follow the Ginzburg-Landau approach used by Toner-Tu (TT) in \cite{toner+tu_95}, namely we identify the underlying symmetries and crucial couplings of the microscopic model, and build the minimal continuous theory compatible with those (for an explicit coarse-graining path see \cite{marchetti_14}). Moreover, it has been shown in \cite{cavagna+al_14} that the spin-overdamped limit of the ISM is the Vicsek model \cite{overdamped}. Because Vicsek model is for TT  hydrodynamic theory what the ISM is for the new theory, we ask that the spin-overdamped limit of our equations must equal to TT.
Using these guidelines, we propose the following dynamical field theory,
\begin{subequations}
\label{SIMhydro}
\begin{align}
  D_t \vv & = \frac1\chi {\vec{s}}\times\vv -\nabla P - \frac{\partial
    V}{\partial \vv}\ , \label{SIMhydro-v}\\
  D_t {\vec{s}} & = \frac{\mathcal{J}}{v_0^2} \vv\times\nabla^2\vv -
  \frac{\eta}{\chi}{\vec{s}} \ , \label{SIMhydro-s}\\
  \partial_t \rho &= -\nabla\cdot(\rho\vv) \ . \label{SIMhydro-rho}
\end{align}
\label{gonzo}
\end{subequations}
Some of the terms in (\ref{gonzo}) are the same as in TT \cite{toner+tu_95}:
$\vv$ and $\rho$ are the velocity and density fields; the material derivative
is defined as $D_t = \partial_t + \lambda \vv\cdot\nabla$, where $\lambda$
breaks the Galilean invariance (we must allow in principle for different
$\lambda$ in the $\vv$ and ${\vec{s}}$ equations);
the pressure $P(\rho)$ is a function of the density; and the confining potential is, $V(\vv) = \int\!\!d^3r\, [ -\alpha/2 \ v^2 +  \beta/2 \ (\vv\cdot\vv)^2 ]$, with $\alpha/\beta = v_0^2$. The novel ingredients coming from the ISM are the spin field, ${\vec{s}}$, its associated inertia, $\chi$, and the spin-velocity coupling, ${\vec{s}}\times\vv$. This cross product indicates that ${\vec{s}}$ is the generator of the rotations of $\vv$, as well as of any other field $f$. This crucial property is expressed by the Poisson relation,
\begin{equation}
\{s, f\}=\frac{d f}{d \varphi} \ ,
\end{equation}
where $s=|{\vec{s}}|$ and $\varphi$ is the phase parametrizing rotations on the plane orthogonal to ${\vec{s}}$ \cite{cavagna+al_14}. Equation (\ref{SIMhydro-s}) is the core of the new theory, because it reinstates inertial effects: the alignment  force, $\mathcal{J}\,\nabla^2\vv$, acts now on, $\dot {\vec{s}}$, rather than on $\dot \vv$, as in Toner-Tu theory. 
Note that this term can be viewed as coming from the same Poisson bracket as the spin-velocity coupling in (1a), provided we augment $V({\bf v})$ with a square-gradient term \cite{attanasi+al_14, cavagna+al_14}. Note also the precise formal resemblance to the coupled dynamics of the direct and staggered magnetizations in the Heisenberg antiferromagnet \cite{halperin_69, mazenko_78}. 
$\mathcal{J}$ is the alignment coupling; in the microscopic model $\mathcal{J}= a^2 J$ , where $a$ is the lattice spacing and $J$ the microscopic (bare) alignment strength \cite{cavagna+al_14}. The damping term, $\eta \, {\vec{s}}$, guarantees that in absence of forces the spin relaxes to zero. Finally, for simplicity we have disregarded diffusion terms of the form $\partial_t f = \Gamma \nabla^2 f$, as their only effect is to renormalize the parameters in (\ref{gonzo}) and have no impact on the dispersion relation.

%%%%%%%%%%%%%%%%%%%
%%%%{\bf The TT and ISM limits.}
%%%%%%%%%%%%%%%%%%%

As usual with viscous dynamics, when the dissipation $\eta$ is very
high compared to the inertia $\chi$, momentum becomes irrelevant and one
obtains the overdamped limit \cite{zwanzig}. 
To study this case one must rescale the time, $t\to \eta^{-1}t$ (and all
other dimensional quantities accordingly - Appendix A),
and take the limit $\chi/\eta^2\to 0$. When this is done, the spin ${\vec{s}}$ can be eliminated, giving,
\begin{subequations}
\label{bogus}
\begin{align}
  D_t \vv & = \mathcal{J}\; \nabla^2\vv -\nabla P -
  \frac{\partial V}{\partial  \vv} \ , \\
  \partial_t \rho &= -\nabla\cdot(\rho\vv) \ .
\end{align}
\end{subequations}
Equations (\ref{bogus})
are the same as TT equations in their simplest form \cite{ramaswamy_review}, with $\mathcal{J}$ playing the role of the 
kinematic viscosity or stiffness depending on whether one views ${\bf v}$ as a velocity or an orientation.
Hence, the spin-overdamped limit of our new field equations gives TT theory, consistently with the fact 
that, in the same limit, the ISM is identical to the Vicsek model \cite{cavagna+al_14}.

%%%%%%%%%%%%%%%%%%%%%%%%%%%%%%%%%%%%
{\it Linear expansion.}
%%%%%%%%%%%%%%%%%%%%%%%%%%%%%%%%%
To check for propagating modes, we study the linear expansion of equations (\ref{gonzo}) in the broken
symmetry phase and in the zero temperature limit (high polarization). 
We consider fluctuations
around the equilibrium values of $\vv$, $\rho$, and
$P$: $  \vv = v_0 \un + \delta\vv$,  $\rho = \rho_0 + \delta\rho$, and
$  P  = P_0 + \sigma \delta\rho$.
 We perform a Galilean transformation to a frame where
the average velocity is zero.  Since $\lambda$ is known to be close to 1 \cite{toner+al_98},
we neglect terms proportional to $\lambda-1$ \cite{ramaswamy_review}.
We introduce the projections of $\delta\vv$ in the directions parallel
and perpendicular to the direction of motion $\un$: $\vpar = (\un\cdot\delta\vv)\un$; 
$\vperp=\delta\vv-\vpar$. In the broken symmetry phase $\vpar$ will be
of higher order than $\vperp$ because the confining potential is flat in
the transverse direction (Goldstone mode).  We can therefore neglect $\vpar$.
Finally, we study the equations in the planar case, in which $\vpar$, $\vperp$ and ${\vec{s}}$ are scalars.  
The linear expansion of equations (\ref{gonzo}) becomes (Appendix B),
\begin{subequations}
  \label{SIMlinear-planar}
  \begin{align}
    \partial_t \vp &= \frac{v_0}{\chi} s - \sigma \partial_\perp \delta\rho ,\\
    \partial_t s &= \frac{\mathcal{J}}{v_0} \nabla^2 \vp - \frac{\eta}{\chi} s, \\
    \partial_t \delta\rho &= -\rho_0 \partial_\perp \vp,
  \end{align}
\end{subequations}
where $\nabla^2 = \partial^2_\parallel + \partial^2_\perp$.
Before studying the existence of propagating modes in
(\ref{SIMlinear-planar}), we consider two limiting cases.

%%%%%%%%%%%%%%%%%
{\it First Sound.}
%\label{sec:massless-limit-1}
The TT limit of overdamping of the spin, $\chi/\eta^2\to 0$, for equations (\ref{SIMlinear-planar}) gives,
\begin{subequations}
  \begin{align}
    \partial_t \vp &= \mathcal{J} \, \nabla^2\vp
    -\sigma \partial_\perp\delta\rho,\\§
    \partial_t \delta\rho &= -\rho_0 \partial_\perp \vp.
  \end{align}
\end{subequations}
We work in polar coordinates in momentum space:  $\theta$ is the angle
between $\kk$ and the direction of motion of the flock (or longitudinal 
direction): $k_\parallel = k\cos\theta$, $k_\perp = k\sin\theta$.
Introducing the speed of {\it first sound}, $c_1^2 \equiv \rho_0\sigma $,
and the damping time, $ \tau_1\equiv 2/k^2\mathcal{J}$,
 the frequencies are given by,
\begin{equation}
  \label{TT}
  \omega_\pm = - i/\tau_1 \pm c_1k \; \sqrt{\sin^2\theta
    - k^2/k_1^2} \ \  ; \ \ k_1 \equiv c_1\tau_1 k^2 \ ,
\end{equation}
which is the dispersion relation of Toner-Tu \cite{toner+tu_98}. Propagating modes requires a nonzero real part of the frequency,
which only happens for $k < k_1\lvert \sin\theta \rvert$.
This has two implications: i) first sound displays {\it anisotropic} propagation ($\theta$ dependence); ii) first sound is overdamped at short wavelengths (large $k$). 
Note that this first sound, unlike that in standard fluids, is a consequence of broken symmetry, 
not of momentum conservation \cite{ramaswamy_review}.

%%%%%%%%%%%%%%%%%%%
{\it Second Sound.}
%%%%%%%%%%%%%%%%%%%%
On the other hand, taking $\sigma\to0$ in~(\ref{SIMlinear-planar}), the spin decouples
from the density, giving,
\begin{subequations}
  \begin{align}
    \partial_t \vp &= \frac{v_0}{\chi} s\ ,\\
    \partial_t s &= \frac{\mathcal{J}}{v_0} \nabla^2\vp - \eta_0 s \ , 
  \end{align}
\end{subequations}
where $\eta_0\equiv \eta/\chi$ is the reduced viscosity.
Introducing the speed of \emph{second sound}, $c_2^2 \equiv \mathcal{J}/\chi$,
and the damping time, $ \tau_2\equiv 2/\eta_0$,
the frequencies can be written as,
\begin{equation}
  \label{SS}
  \omega_\pm = -i/\tau_2 \pm c_2 k \; \sqrt{1- k_2^2/k^2} \ ; \ k_2\equiv 1/c_2\tau_2  \ .
\end{equation}
This dispersion relation has been obtained in \cite{attanasi+al_14} and \cite{cavagna+al_14} under the approximation that the time scale of collective turns is so short that the network is almost fixed, so that density fluctuations can be neglected.
Unlike the first sound mode of TT, which travels over density fluctuations, the mode in (\ref{SS}) describes a density-independent spin-wave, that is a propagating disturbance purely of the orientations; this mode would propagate also on a fixed lattice, with {\it zero} density fluctuations. In analogy with the spin-wave theory of superfluidity \cite{halperin_69}, we call this mode `second sound' \cite{attanasi+al_14}. There are two fundamental differences between second sound (\ref{SS}) and first sound (\ref{TT}): i) the dispersion relation for second sound is \emph{isotropic,} in particular, second sound can propagate also in the parallel direction, whereas first sound cannot; ii) $\omega$ has a real part only for $k > k_2$, hence second sound is overdamped at long wavelengths (small $k$).

%%%%%%%%%%%%%%%%%%%%%%%%%%%%%%%%%%%%%%%%%%%%%%%%%%%%%%%%%
\begin{figure}
  \includegraphics[width=7 truecm]{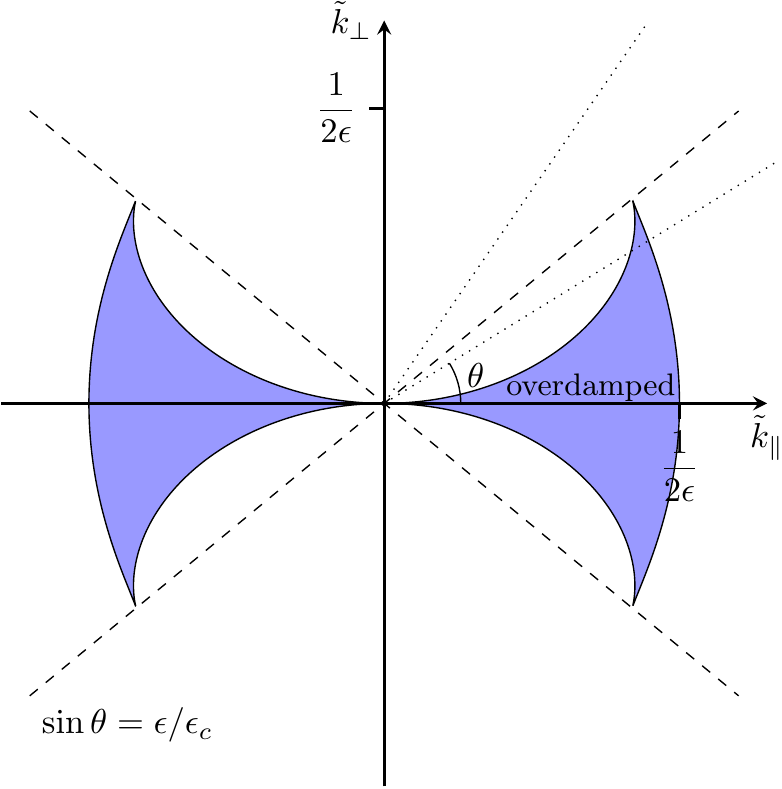}
  \caption{{\bf Small speed ratio: $\epsilon<\epsilon_c$.} The overdamped, nonpropagating  zone (blue) has purely imaginary frequency,
  while in the propagating zone (white) $\mathrm{Re}\; \tilde\omega\neq 0$. 
     The dotted lines indicate two paths along which $k$ is changed. For $\sin\theta < \epsilon/\epsilon_c$
    a gap crossing the nonpropagating region emerges. 
    }
  \label{fig:overdamped-region-2}
\end{figure}
%%%%%%%%%%%%%%%%%%%%%%%%%%%%%%%%%%%%%%%%%%%%%%%%%%%%%%%%%

%{\bf Full dispersion relation.}
%\label{sec:full-planar-linear}
To understand the crossover between first and second sound we must study the full linearized equations (\ref{SIMlinear-planar}). We have three fields (velocity, spin and density), 
and therefore three frequency modes, given by the solutions of the following dispersion relation (Appendix C),
\begin{equation}
  \omega^3 + i\eta_0 \omega^2 - (c_1^2k_\perp^2 + c_2^2k^2)\omega -
  i\eta_0 c_1^2 k_\perp^2 = 0 \ .
  \label{disp-rel-eq-full}
\end{equation}
It is convenient to introduce the dimensionless frequency, $\tilde\omega\equiv \omega/\eta_0$, and 
the dimensionless momentum, $\tilde \kk \equiv c_1 \kk /\eta_0$. Once this is done, we find that the 
dispersion relation only depends on one key parameter, $\epsilon\equiv c_2/c_1$, 
\begin{equation}
\tilde  \omega^3 + i \tilde \omega^2 - \tilde k^2(\sin^2\theta + \epsilon^2)\,\tilde\omega -
  i \tilde k^2\sin^2\theta= 0 \ .
  \label{nuke}
\end{equation}
The parameter $\epsilon$ is the second-to-first sound speed ratio; much of the 
propagation properties of the new theory depend on the speed ratio, namely on how fast is second sound compared to first.  After a little algebra (Appendix C), one finds that there is a critical value of the speed ratio, $\epsilon_c = \sqrt{8}$, 
separating two very different
regimes.

%%%%%%%%%%%%%%%%%%%%%%%%%%%%%%%%%%%%%%%%%%%%%%%%%%%%%%%%%
\begin{figure}
  \includegraphics[width=7 truecm]{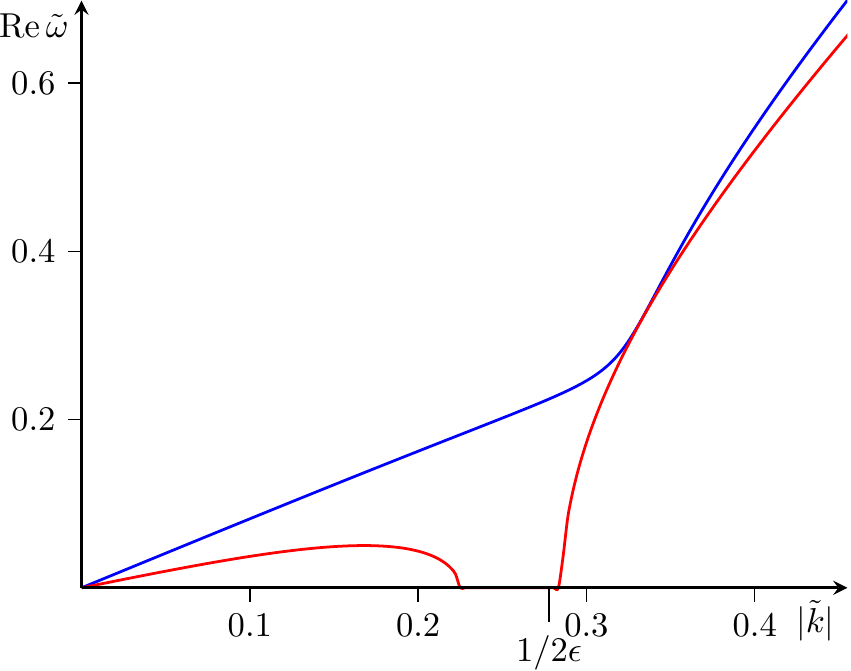}
  \caption{{\bf Small speed ratio: $\epsilon<\epsilon_c$.}
  Real part of the frequency vs.\ $\lvert \tilde k \rvert$ along
   two different directions (dotted lines of Fig.1). For $\sin\theta > \epsilon/\epsilon_c$ (blue) there is no gap and all
   wavelengths are propagating. On the other hand, for $\sin\theta < \epsilon/\epsilon_c$ (red) an overdamped
   gap emerges, separating first sound - low $\tilde k$ -  from second sound - large $\tilde k$.
    }
  \label{fig:omegavsk-three-paths}
\end{figure}
%%%%%%%%%%%%%%%%%%%%%%%%%%%%%%%%%%%%%%%%%%%%%%%%%%%%%%%%%

%%%%%%%%%%%%%%%%
% $\epsilon  <  \epsilon_c$
%%%%%%%%%%%%%%%%

{\it Small speed ratio: $\epsilon<\epsilon_c$.}
When second sound speed is not too large compared to  first sound, we have the situation depicted in Fig.1: there is a region  along
the longitudinal momentum axis where the real part of the frequency is zero for all modes, 
so that no propagation can take place.  In the rest of the $\vec k$ plane two modes (out of three) have
$\mathrm{Re}\,\omega(k)\neq 0$, so that propagation occurs. Let us fix a direction $\theta$ of
the wave vector $\vec k$ and follow a path by increasing the modulus $k$. If $\theta$ is large enough ($\sin\theta
> \epsilon/\epsilon_c$),  the real part of $\omega(k)$ is always nonzero, so that there is
propagation at all wavelengths  (Fig.2). On the other hand, if $\theta$ is small
($\sin\theta < \epsilon/\epsilon_c$) the path crosses the overdamped region, and a {\it gap} emerges: 
the propagating regimes for small and large
$k$ are separated from each other by a nonpropagating region at intermediate $k$ (Fig.2).
This gap separates the first sound region at low $k$, from the second sound region at large $k$
(see Appendix D for the $k\to 0$ vs $k\to\infty$ exact solutions). For large $\theta$ there is 
hybridization of first and second sound and the crossover from one mode to the other is smooth.

%%%%%%%%%%%%%%%%%%%%%%%%%%%%%%%%%%%%%%%%%%%%%%%%%%%%%%%%%
\begin{figure}[ht]
  \includegraphics[width=7 truecm]{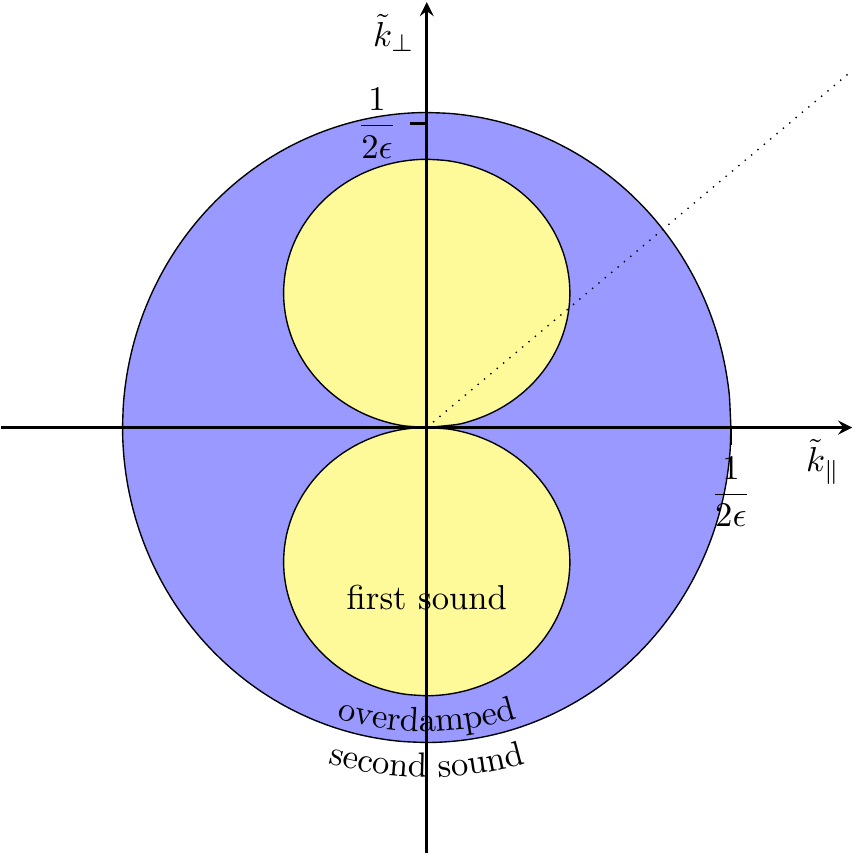}
  \caption{{\bf Large speed ratio: $\epsilon>\epsilon_c$.} 
 The two propagating zones (nonzero real part of the frequency) are now separated by an overdamped ring (blue).
Any path in momentum space must cross the overdamped ring, giving rise to a gap.}
  \label{fig:overdamped-region-1}
\end{figure}
%%%%%%%%%%%%%%%%%%%%%%%%%%%%%%%%%%%%%%%%%%%%%%%%%%%%%%%%%

%%%%%%%%%%%%%%%%
% $\epsilon  >  \epsilon_c$
%%%%%%%%%%%%%%%%
{\it Large speed ratio: $\epsilon>\epsilon_c$.}
When $\epsilon$ increases the nonpropagating region grows in size: the tips of the left and right wedges  approach
each other and eventually touch for $\epsilon = \epsilon_c$, sealing the first sound pocket at small $k$. 
For $\epsilon  >  \epsilon_c$, the situation is the one shown in Fig.3: the two propagating regions
 are now completely disconnected. 
This means that when we plot the real part of the frequency as a function of $k$, 
we are bound to find a nonpropagating gap between first and second sound, {\it no matter what is the path we follow 
in momentum space} (Fig.4). 
%%%%%%%%%%%%%%%%%%%%
% Physical significance of the gap
%%%%%%%%%%%%%%%%%%%%%
When the second sound speed $c_2$ is much larger than the first sound speed $c_1$
($\epsilon > \epsilon_c$), spin fluctuations propagate much faster than density fluctuations. 
Hence, in the $\epsilon > \epsilon_c$ regime the turning information propagates
on a much shorter timescale than density fluctuations and assuming a 
fixed network during spin propagation is justified. This is the physical meaning of the separation between
first and second sound for $\epsilon > \epsilon_c$ and the reason why experimental data on turning flocks are in agreement with 
the fixed network approximation of \cite{attanasi+al_14}.

%%%%%%%%%%%%%%%%%%%%%%%%%%%%%%%%%%%%%%%%%%%%%%%%%%%%%%%%%
\begin{figure}[ht]
  \includegraphics[width=7 truecm]{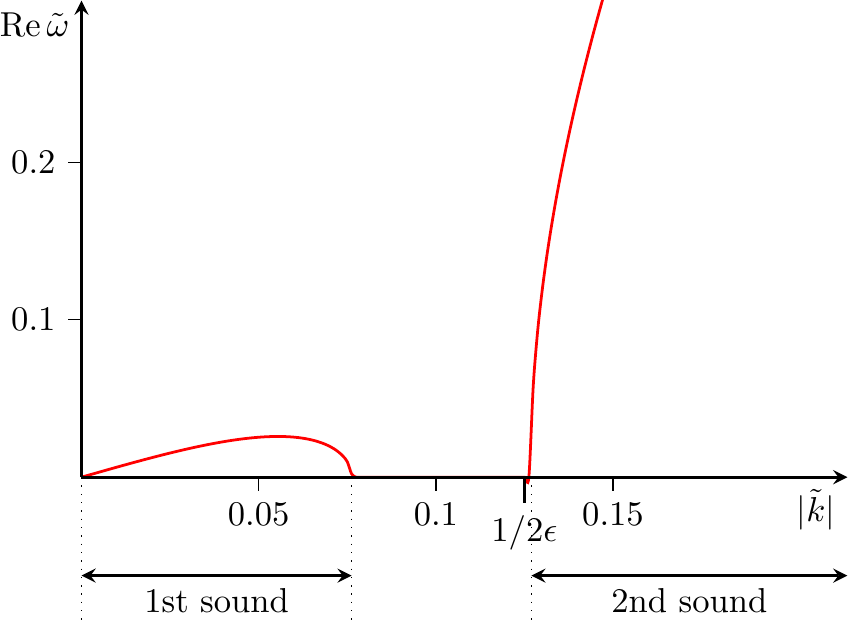}
  \caption{
  {\bf Large speed ratio: $\epsilon>\epsilon_c$.}
   Real part of the frequency vs.\ $\lvert \tilde k \rvert$ at one given value of $\theta$ (dotted line in Fig.3). A nonpropagating gap
    between first and second sound is found at any $\theta$. For $\theta=0$ the gap extends down 
    to $k=0$. The lower band edge of second sound occurs at $\tilde k \sim 1/2\epsilon$,
    that is $k=\eta_0/2c_2 = k_2$, consistent with the second sound limit (\ref{SS}).
}
  \label{fig:omegavsk-real}
\end{figure}
%%%%%%%%%%%%%%%%%%%%%%%%%%%%%%%%%%%%%%%%%%%%%%%%%%%%%%%%%

%%%%%%%%%%%%%%%%%%%%%%%%%
%\section{Propagation across a finite system}
%\label{finite_size}
%%%%%%%%%%%%%%%%%%%%%%%%%

{\it Finite size.} 
In a system of finite size $L$, a nonzero real part of the frequency is not enough to grant information transfer.
Modes are damped with a characteristic time $\tau(k)=1/\mathrm{Im}\,\omega(k)$ and cross-system propagation only occurs when the distance traveled by the signal before damping is larger than $L$. By  dimensional analysis, this distance is $c_1\tau_1 \sim k_1/k^2$ for first sound, and $c_2\tau_2 \sim  1/k_2$ for second sound (eqs. (\ref{TT}) and (\ref{SS})). Moreover,  the maximum wavelength traveling through the system cannot exceed its size.  These two conditions give,
\begin{equation}
1/k < L < k_1/k^2 \quad ; \quad 1/k < L < 1/k_2 \ ,
\end{equation}
for first and second sound, respectively. By collapsing the two sides of each inequality, we obtain that cross-system propagation can only occur if $L>1/k_1$ for first sound, and if $L<1/k_2$ for second sound. Therefore, if there is a gap in momentum space, namely if $k_1<k_2$, we obtain a corresponding gap in $L$ for $1/k_1 <L<1/k_2$. We conclude that there is a regime of medium sized flocks that are `silent': {\it no signal can cross the system at any wavelength}.  Using the full dispersion relation, and therefore the true phase velocity $c(k)$ and damping time $\tau(k)$, does not change qualitatively the dimensional argument (Appendix G). In fact, it is possible to show that the gap in $L$  appears even {\it before} the gap in $k$, in the region where $\mathrm{Re\ }\omega(k)$ has a minimum. 

Second sound is essential to transfer directional information across natural flocks \cite{attanasi+al_14}. The fact that second sound is damped in large systems may be responsible for an upper cutoff in the size of flocks performing collective turns. Very large flocks exist, but they may have troubles to collectively change direction of motion. Even though we have no data on huge flocks, i.e. of the order $10^4$ individuals or larger, our anecdotal experience in the field agrees with this conclusion. Second, at the more speculative level, the existence of a silent regime at intermediate sizes, where no information what-so-ever can propagate (not spin, nor density fluctuations), suggests that size control may be a more general concern of biological groups than previously thought.

%\vskip 0.1 truecm

%%%%%%%%%%%%%%%%%%%%%%%%%%%%%%%%
% acknowledgments
%%%%%%%%%%%%%%%%%%%%%%%%%%%%%%%%

AC acknowledges US-AFOSR grant FA95501010250 (through the University of Maryland), 
IG grants IIT--Seed Artswarm and ERC--StG n.257126.
SR acknowledges a J C Bose Fellowship, the hospitality of the Initiative for
Theoretical Sciences at the CUNY Graduate Center, the support of the Kavli
Institute for Theoretical Physics under Grant No. NSF PHY11-25915, and
discussions with M C Marchetti and X Yang.

%%%%%%%%%%%%%%%%%%%%%%%%%%%%%%%%%%%%%%%%%%%%%%%%%%%%%%%%%
%%%%%%%%%%%%%%%%%%%%%%%%%%%%%%%%%%%%%%%%%%%%%%%%%%%%%%%%%
%%%%%%%%%%%%%%%%%%%%%%%%%%%%%%%%%%%%%%%%%%%%%%%%%%%%%%%%%

\appendix

%\section*{Supplemental Information}

\section{Rescaling in the overdamped limit}

To study the spin-overdamped limit one must rescale the time
$t\to \eta^{-1}t$, which implies the substitutions $\vv\to
\eta^{-1}\vv$, ${\vec{s}}\to \eta^{-2}{\vec{s}}$, $\alpha\to \eta^{-1}\alpha$,
$\beta\to \eta\beta$, $P\to \eta^{-2} P$.  After taking
$\eta\to\infty$ the $D_t{\vec{s}}$ term disappears and ${\vec{s}}$ can be
eliminated from the equations, giving
\begin{subequations}
\label{massless-hydro}
\begin{align}
  D_t \vv & = - \frac{\mathcal{J}}{v_0^2} \vv\times
  \vv\times\nabla^2\vv -\nabla P -
  \frac{\partial V}{\partial  \vv}, \\
  \partial_t \rho &= -\nabla\cdot(\rho\vv).
\end{align}
\end{subequations}
The double cross product can be expanded as
\begin{equation}
  \label{eq:1}
  - \frac{\mathcal{J}}{v_0^2} \vv\times \vv\times\nabla^2\vv = 
 \frac{\mathcal{J} v^2}{v_0^2} \nabla^2\vv -\frac{\mathcal{J}}{v_0^2}
 \vv(\vv\cdot\nabla^2\vv),
\end{equation}
which at first order equals $\mathcal{J} \nabla^2\vv$.  At higher orders,
the second term can be ignored anyway, since its effect in principle
is to cancel the parallel projection of the Laplacian, giving no
parallel projection of $D_t\vv$ and hence constant modulus of $\vv$.
However this is useless in the presence of the $\nabla P$ term, which
can have a nonzero parallel projection.  It is then the job of the
potential to fix (with fluctuations) the value of $\lvert\vv\rvert$.

\section{Linear expansion of the equations}

We are interested in the strongly ordered (broken symmetry, high
polarization) limit of Eqs.~(\ref{gonzo}).  We thus expand $\vv$, $\rho$,
and $P$, and keep only first order terms in $\delta\vv$, $\delta\rho$,
and $\delta\rho$.  Then a Galilean transformation to a frame where the
average velocity is zero is done.  After neglecting terms proportional
to $\lambda-1$ (since $\lambda$ is known to be close to 1
\cite{toner+al_98}), we obtain
\begin{subequations}
  \label{SIMlinear}
  \begin{align}
    \partial_t \delta\vv &= \frac{1}{\chi}{\vec{s}}\times v_o\un - \sigma
    \nabla \delta\rho - 2\beta v_0^2 (\un\cdot\delta\vv)\un,\\
    \partial_t {\vec{s}} &= \frac{\mathcal{J}}{v_0} \un\times\nabla^2 \delta\vv -
    \frac{\eta}{\chi} {\vec{s}}, \\
    \partial_t \delta\rho &= -\rho_0 \nabla\cdot\delta\vv,
  \end{align}
\end{subequations}
where $\un$ is the direction of the polarization (average $\vv$).
Introducing the projections of $\delta\vv$, $\vpar =
(\un\cdot\delta\vv)\un$, $\vperp=\delta\vv-\vpar$, we can further
neglect $\vpar$, since in the broken symmetry phase it be of higher
order than $\vperp$ because the confining potential is flat in the
direction of $\vperp$.  We thus arrive at the linear system
\begin{subequations}
  \label{SIMlinear-nopar}
  \begin{align}
    \partial_t \vperp &= \frac{1}{\chi}{\vec{s}}\times v_o\un - \sigma
    \nabla_\perp \delta\rho ,\\
    \partial_t {\vec{s}} &= \frac{\mathcal{J}}{v_0} \un\times\nabla^2 \vperp -
    \frac{\eta}{\chi} {\vec{s}}, \\
    \partial_t \delta\rho &= -\rho_0 \nabla_\perp\cdot\vperp.
  \end{align}
\end{subequations}
For simplicity in the paper we have focused on the \emph{planar}
version of Eqs.~\ref{SIMlinear-nopar}, i.e.\ the case where the
direction of ${\vec{s}}$ is independent of position $\rr$.  Since, ${\vec{s}}$ is
perpendicular to $\vv$ \cite{cavagna+al_14}, this means that the velocity
moves in a plane, and thus $\vpar$, $\vperp$ and ${\vec{s}}$ are scalars and
Eqs.~\ref{SIMlinear-nopar} reduce to Eqs.~(\ref{SIMlinear-planar}) of the main
text.  This case is realized when the initial condition is such that
the field $\vv$ lies in the same plane at every point in space.

%%%%%%%%%%%%%%%%%%%%%%%%%%%%%%%%%%%%%%%%%%%%%%%%%%%%%%%%%%%%%%%%%%%%%%%%%%%%%%%

\section{Derivation and solution of the dispersion relation}
\label{dispersion}

After a Fourier transform, Eqs.~(\ref{SIMlinear-planar}) reduce to the
eigenvalue problem
\begin{equation}
  \label{eq:4}
  M\ww = -i\omega \ww,
\end{equation}
where
\begin{equation}
  \label{eq:3}
  \ww = \begin{pmatrix} \vp \\\delta\rho \\ s   \end{pmatrix},
 \qquad
  M = \begin{pmatrix} 0 & -i\sigma k_\perp & v_0/\chi \\
                      -i\rho_0 k_\perp & 0 & 0\\    
                      -\frac{\mathcal{J}}{v_o} k^2 & 0 & -\eta/\chi  \end{pmatrix}.
\end{equation}
The allowed frequencies are found by solving
$\det(M+i\omega\mathbb{1})=0$, which gives,
\begin{equation}
  \omega^3 + i\eta_0 \omega^2 - (c_1^2k_\perp^2 + c_2^2k^2)\omega -
  i\eta_0 c_1^2 k_\perp^2 = 0. \label{disp-rel-eq-full}
\end{equation}
Introducing polar coordinates ($k_\perp=k\sin\theta$) and the
adimensional quantities
\begin{equation}
   \epsilon = \frac{c_2}{c_1}, \qquad \tilde
   \omega=\frac{\omega}{\eta_0}, \qquad
   \tilde k = \frac{c_1 k}{\eta_0} , 
\end{equation}
this is Eq.(\ref{nuke}) of the main text.  It is convenient to define a
new variable $u$ through $\tilde \omega=iu$, because the equation for
$u$ has real coefficients:
\begin{equation}
  u^3 + u^2 + {\tilde k}^2(\sin^2\theta + \epsilon^2) u +
  \sin^2\theta {\tilde k}^2 = 0. \label{disp-rel-eq-real}
\end{equation}
This equation must then have at least one real root, i.e.\ one of the
modes will always be overdamped (purely imaginary $\omega$).
Propagating (but damped) modes will exist when
(\ref{disp-rel-eq-real}) has complex roots.  This is controlled by the
discriminant
\begin{equation}
  \label{eq:8}
  \begin{split}
  \Delta(\tilde k,\sin\theta) = {}& -4{\tilde k}^2\sin^2\theta
    -4{\tilde k}^6(\epsilon^2+\sin^2\theta)^3 + \\
   & {\tilde k}^4 \left(\epsilon^4 + 20\epsilon^2\sin^2\theta -
     8\sin^4\theta \right).
  \end{split}
\end{equation}
When $\Delta\ge 0$, all modes are overdamped (all roots of
(\ref{disp-rel-eq-real}) are real).  It is convenient to study instead
the sign of $\tilde\Delta=\Delta/{\tilde k}^6$, writing it as a
function of $y={\tilde k}^{-2}$,
\begin{equation}
  \label{eq:9}
  \begin{split}
  \tilde \Delta(y,\sin\theta) ={}& -4 y^2 \sin^2\theta  - 4 (\epsilon^2+\sin^2\theta)^3 + \\
&  y \,  (\epsilon^4 + 20 \epsilon^2\sin^2\theta -8\sin^4\theta)  \ .
  \end{split}
\end{equation}
Since the coefficient of the quadratic term in $y$ is negative,
$\tilde \Delta$ (and hence $\Delta$) will be positive between the
roots $y_\pm$, provided they are real.  The roots will be real
whenever
\begin{equation}
  \label{eq:100}
  D_2(\epsilon,\theta) = \epsilon^2 (\epsilon^2 - 8\sin^2\theta)^3 \ge 0.
\end{equation}
It turns out that if the roots are real, they are also positive, so
that all that is required for the existence of an overdamped interval
as a function of $\tilde k$ for fixed $\theta$ is that
\begin{equation}
  \epsilon^2 >8\sin^2\theta \ .
  \label{nupe}
\end{equation}
The largest value of the right hand side of (\ref{nupe}) this gives the critical value
used in the main text,
\begin{equation}
\epsilon_c=\sqrt{8} \ .
\end{equation}

%%%%%%%%%%%%%%%%%%%%%%%%%%%%%%%%%%%%%%%%%%%%%%%%%%%%%%%%%%%%%%%%%%%%%%%%%%%%%%%

\section{Asymptotic behavior of the frequency}

\begin{figure}
\includegraphics[width=0.9\columnwidth]{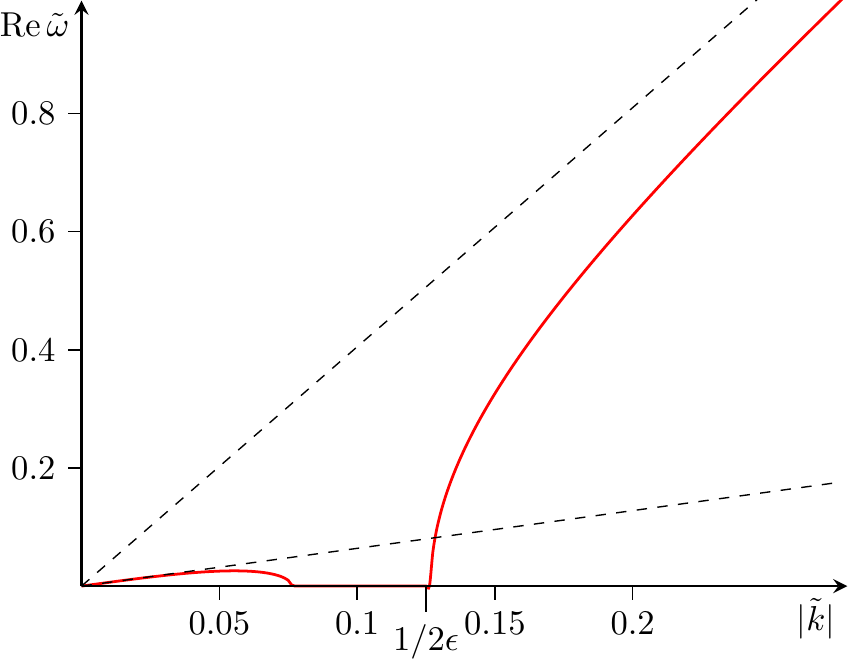}
  \caption{{\bf Comparison with the asymptotic expansion.} Dispersion relation for $\epsilon=4>\epsilon_c$ and
    $\theta=40^\circ$ (full curve) showing the asymptotic lines
    (dashed) for $\tilde k\to0$ and $\tilde k\to\infty$.}
  \label{fig:omega-asympt}
\end{figure}

Equation~(\ref{disp-rel-eq-full}) can be solved in the $k\to0$ and
$k\to\infty$ limits.  For $k\to0$ we find
\begin{subequations}
\label{eq:7}
\begin{align}
  \omega_\pm & = \pm c_1 k \sin\theta - i\frac{\mathcal{J}}{2}k^2  + O(k^3),\\
  \omega_3 & = -i\eta_0 - i\mathcal{J} k^2 + O(k^3).
\end{align}
\end{subequations}
In this limit, Eqs.~(\ref{eq:7}) give a purely imaginary mode plus two
modes that coincide with the first sound (TT) modes
(Eq.~\ref{TT}).  For $k\to\infty$ we have instead
\begin{subequations}
\label{eq:16}
\begin{align}
\omega_\pm & = -i\frac{\eta_0}{2}
  \frac{1}{1+\frac{\sin^2\theta}{\epsilon^2}} \pm 
   c_2 k \sqrt{1+\frac{\sin^2\theta}{\epsilon^2}} + O(1/k),\\
  \omega_3 & = -i \eta_0 \frac{\sin^2\theta}{\epsilon^2+\sin^2\theta} + O(1/k).
\end{align}
\end{subequations}
Eqs.~\ref{eq:16} yield again a purely imaginary mode plus an
anisotropic version of the ISM dispersion law (second sound,
Eq.~(\ref{SS})).  The ISM law is recovered at first order for
$\epsilon\to\infty$ ($c_2\gg c_1$), and at all orders for $\theta=0$
as it is easy to check solving directly Eq.~\ref{disp-rel-eq-full} in
this case.  This analysis confirms our identification of the modes as
the TT modes for $k\to0$ and ISM modes for $k\to\infty$.
Fig.~\ref{fig:omega-asympt} shows the dispersion relation for the
propagating modes together with the asymptotic solution.

\begin{figure}
\includegraphics[width=0.9\columnwidth]{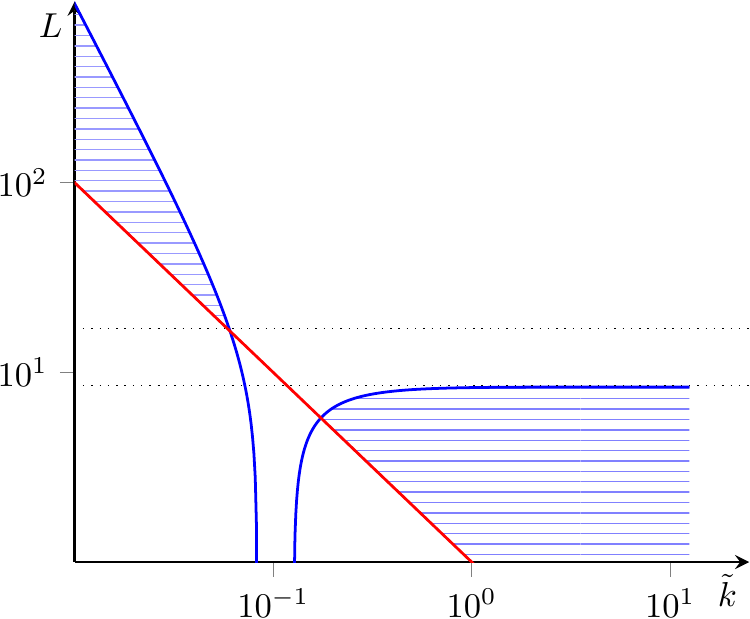}
  \caption{{\bf Silent flocks}. Propagation diagram in the $(L,\tilde k)$ plane.
  Blue line: the upper bound of inequality (\ref{eq:18}), namely the maximum distance, $l_\mathrm{max}(k)$, traveled by a monochromatic wave before damping. Red line:  the lower bound of (\ref{eq:18}), $1/k$. The shaded regions are those satisfying inequality (\ref{eq:18}), where cross-system propagation is possible.
  }
  \label{umpa}
\end{figure}
%%%%%%%%%%%%%%%%%%%%%%%%%%%%%%%%%%%%%%%%%%%%%%%%%%%%%%%%%%%%%%%%%%%%%%%%%%%%%%%

\section{Checking for instabilities}

The imaginary part of the frequency indicates a non oscillating
exponential prefactor multiplying the plane wave solutions.  With our
sign choice for the Fourier transforms, $\mathrm{Im}\ \omega<0$ means an
exponentially \emph{damped} wave, while $\mathrm{Im}\ \omega>0$ would give an
exponentially growing prefactor, implying an instability in the
solution (i.e.\ a perturbation that moves arbitrarily far away from
the reference state, in our case the highly polarized flock).

As Eqs.~\ref{eq:7} and~\ref{eq:16} show, the asymptotic regimes are
stable.  We thus ask if $\mathrm{Im}\ \omega$ changes sign at some finite $k$.
Again it is more convenient to consider the equation for $u$
(Eq.~(\ref{disp-rel-eq-real}), where now we are interested in the real
part ($\mathrm{Im}\ \omega=\mathrm{Re}\ u$).  Since all coefficients in
Eq.~(\ref{disp-rel-eq-real}) are positive, it is clear that when the
solution $u$ is real, it must be negative.  Consider then the case
where the solution is $u=u'+iu''$ with $u''\neq0$.  Looking for a
change in the sign of $u'$, assume $u'=0$ and write the real and
imaginary parts of Eq.~(\ref{disp-rel-eq-real}):
\begin{align}
  -u''+(\sin^2\theta+\epsilon^2){\tilde k}^2 &=0,\\
  -u''^2+\sin^2\theta{\tilde k}^2 &=0.
\end{align}
For $\epsilon\neq0$, these equations can only be satisfied for $\tilde
k=0$, i.e.\ $u'$ (and thus $\mathrm{Im}\ \omega$) cannot change sign for
$k\in(0,\infty)$.

Of course, this analysis is limited to the present formulation of the
hydrodynamic theory.  As discussed above, outside the $k\to0$ limit
the applicable theory is not necessarily universal, and alternative
hydrodynamic formulations are possible.  Ultimately,
discriminating among alternative theories will need the comparison with 
experimental data.

%%%%%%%%%%%%%%%%%%%%%%%%%%%%%%%%%%%%%%%%%%%%%%%%%%%%%%%%%%%%%%%%%%%%%

\section{Viscosity gap}
\label{gap_in _viscosity}

%%%%%%%%%%%%%%%%%%%%%%
% momentum -viscosity duality
%%%%%%%%%%%%%%%%%%%%%%

The dispersion relation (\ref{nuke}) depends on the dimensionless momentum $c_1 k/\eta_0$. 
Hence, there is a duality between the low $k$ regime and the large $\eta_0$ regime
(and vice-versa). This means that the gap we have found in 
momentum space can also be viewed as a gap in spin viscosity (at fixed $k$): for low $\eta_0$
the frequency has a nonzero real part and the system displays second sound propagation; as we increase the viscosity, spin fluctuations get damped and we eventually enter into the gap, where every mode is purely imaginary; but then,
by further {\it increasing} the spin viscosity, we exit the gap and enter the regime of first sound propagation, where spin inertia is irrelevant and density fluctuations are the carrier of the signal. 

This counterintuitive phenomenon is due to the fact that the viscosity $\eta$ is
a {\it spin} viscosity, so that it damps spin fluctuations, not density
fluctuations. For this same reason, the overdamped limit of the theory (the TT
limit) is a regime of spin overdamping, not of general overdamping, as clearly
shown by the fact that first sound propagation of density fluctuations emerges
deeply into the spin-overdamped region \cite{overdamped}.

%%%%%%%%%%%%%%%%%%%%%%%%%%%%%%%%%%%%%%%%%%%%%%%%%%%%%%%%%%%%%%%%%%%%%

\section{Propagation across a finite system}
\label{finite_size}

We have discussed the existence of regimes with propagating modes.
However, even if the real part of the frequency is nonzero,
its imaginary part is nonzero as well, due to presence of
spin dissipation $\eta_0$, so that even propagating modes are damped
with a characteristic time $\tau(k)=1/\mathrm{Im\,}\omega(k)$.  A monochromatic wave
travels with a (phase) velocity $c(k)=\mathrm{Re\,}\omega(k)/k$, so that the distance
it can travel without significant damping is roughly.
\begin{equation}
  \label{eq:17}
  l_\mathrm{max}(k) \simeq c(k)\tau(k) = \frac{\mathrm{Re\,}\omega(k)}{k\,\mathrm{Im\,}\omega(k)} \ .
\end{equation}
For a finite system of size $L$, we can then expect that a signal will
travel across the whole system when $l_\mathrm{max}(k) > L$.
Boundary conditions will typically impose a lower bound on $k$ of the
order of $1/L$, so the waves are expected to propagate across the
system for values of $k$ and $L$ satisfying the inequality,
\begin{equation}
 \label{eq:18}
   1/k < L <   l_\mathrm{max}(k)  \ .
\end{equation}
The plot of $l_\mathrm{max}(k)$ is shown in Fig.\ref{umpa}, together with $1/k$. For $\epsilon > \epsilon_c$ the gap in $k$ implies an equivalent gap in $L$, where neither first nor second sound can propagate (silent flocks). The results of the dimensional analysis performed in the main text are therefore qualitatively correct. As we have shown in the main text, the second sound band (low $k$) gets reduced for decreasing $\theta$ (for $\theta=0$ the gap extends down to $k=0$ and there is no first sound). For this reason, the gap in $k$, and therefore that in $L$, are larger the lower $\theta$.

Finally, it is interesting to note that  the gap in $k$ is not strictly necessary to have a gap in $L$. Right before being gapped, the curve $l_\mathrm{max}(k)$ has a minimum in the region of $k$ where the gap is about to open. This is sufficient to produce a gap in $L$.

%%%%%%%%%%%%%%%%%%%%%%%%%%%%%%%%%%%%%%%%%%%%%%%%%%%%%%%%%
%%%%%%%%%%%%%%%%%%%%%%%%%%%%%%%%%%%%%%%%%%%%%%%%%%%%%%%%%
%%%%%%%%%%%%%%%%%%%%%%%%%%%%%%%%%%%%%%%%%%%%%%%%%%%%%%%%%
%%%%%%%%%%%%%%%%%%%%%%%%%%%%%%%%%%%%%%%%%%%%%%%%%%%%%%%%%
%%%%%%%%%%%%%%%%%%%%%%%%%%%%%%%%%%%%%%%%%%%%%%%%%%%%%%%%%


\begin{thebibliography}{99}
\bibitem{overdamped}
The theory we are about to introduce gives propagating modes even in the overdamped limit. This odd behaviour is due to the fact that the
infinite spin viscosity limit of the ISM (a limit normally called `overdamped') reproduces the Vicsek model, which has the propagating modes of
Toner-Tu. To avoid this ambiguity, we have decided to term the large $\eta$ limit, 
 `spin-overdamped', rather than simply `overdamped'.
\end{thebibliography}
\end{document}